\documentclass[11pt,a4paper]{article}
\usepackage{latexsym,makeidx,syntonly,amsfonts,amssymb}
\newcommand {\cA}{{\cal A}}
\newcommand {\cB}{{\cal B}}
\newcommand {\cC}{{\cal C}}

\newcommand {\ZBZrz}{{{Z_0}^{(r)}\zbz,\bar{{Z_0}}^{(r)}\zbz}}
\newcommand {\Zrz}{{{Z_0}^{(r)}\zbz}}

\newcommand {\cK}{{\cal K}}

\newcommand {\cN}{{\cal N}}

\newcommand {\sS}{{\cal S}}
\newcommand {\cT}{{\cal T}}
\newcommand {\cS}{{\cal S}}

\newcommand {\cV}{{\cal V}}
\newcommand {\cW}{{\cal W}}
\newcommand {\cX}{{\cal X}}

\newcommand{\zbz}{(z,\bz)}

\newcommand{\Zr}{Z^{(r)}} 

\newcommand{\Ch}{{C^{(2)}}}

\newcommand{\lra}{ \longrightarrow }

\def\j{\psi}
\def\k{\kappa}

\newcommand{\ovl}[1]{\overline{#1}} 
\newcommand{\bz}{\ovl{z}}

\newcommand{\prt}{\partial}

\newcommand{\wbw}{(w,\ovl{w})}  
\newcommand{\sect}[1]{\setcounter{equation}{0}\section{\boldmath#1}}

\newcommand{\be}{\begin{equation}}
\newcommand{\ee}{\end{equation}}
\newcommand{\bea}{\begin{eqnarray}}
\newcommand{\eea}{\end{eqnarray}}

\newcommand{\lbl}[1]{\label{eq:#1}}
\newcommand{\rf}[1]{(\ref{eq:#1})}
\newcommand{\nn}{\nonumber}
 
\newcommand{\beq}{\begin{equation}}
\newcommand{\eeq}{\end{equation}}

\def\endtitle{\end{quotation}\newpage}     
 
\newtheorem{theorem}{Theorem}  
 
\newtheorem{Ansatz}{Statement}
\begin{document}

\pagestyle{empty}


\font\fifteen=cmbx10 at 15pt
\font\twelve=cmbx10 at 12pt

\begin{titlepage}

\begin{center}
\renewcommand{\thefootnote}{\fnsymbol{footnote}}

{\twelve Centre de Physique Th\'eorique\footnote{
Unit\'e Propre de Recherche 7061
}, CNRS Luminy, Case 907}

{\twelve F-13288 Marseille -- Cedex 9}

\vskip 2.5cm

{\fifteen  Primary currents and  Riemannian geometry in $\cW$-algebras} 

\vskip 1.5cm

\setcounter{footnote}{0}
\renewcommand{\thefootnote}{\arabic{footnote}}

{\bf G. BANDELLONI} $^a$\footnote{e-mail : {\tt beppe@genova.infn.it}}
\hskip .2mm  and \hskip .7mm {\bf S. LAZZARINI} $^b$\footnote{and also
Universit\'e de la M\'editerran\'ee, Aix-Marseille II. e-mail : {\tt
sel@cpt.univ-mrs.fr} }\\[6mm] 
$^a$ \textit{Dipartimento di Fisica
dell'Universit\`a di Genova,}\\
{\it Via Dodecaneso 33, I-16146 GENOVA, Italy}\\
 and \\
{\it Istituto Nazionale di Fisica Nucleare, INFN, Sezione di Genova}\\
{\it via Dodecaneso 33, I-16146 GENOVA, Italy}\\[4mm]
$^b$ {\it Centre de Physique Th\'eorique, CNRS Luminy, Case 907,}\\
{\it F-13288 MARSEILLE Cedex, France} 
\end{center}

\vskip 1.cm

\centerline{\bf Abstract}

It is proved that general consistency requirements of stability under
complex analytic change of charts show that primary currents in finite chiral
$\cW$-algebras are described in terms of pure gravitational variables.

\vskip 2.5cm

\noindent 1998 PACS Classification: 11.10 - 11.25 hf

\indent


\indent

\noindent{CPT--2000/P.4206}, Internet : {\tt www.cpt.univ-mrs.fr}

\end{titlepage}

\renewcommand{\thefootnote}{\arabic{footnote}}
\setcounter{footnote}{0}
\pagestyle{plain}
\setcounter{page}{1}

\newpage

\section{Introduction}

The description of a physical phenomenon is independent of any
parametrization used for its representation.  Without this belief  any
physical law loses its credibility.

This degree of freedom helps us in many complex sets of circumstances
 where symmetries or  work simplification conditions are present.
 Mathematically  speaking, any switch of parametrization  is realized
 by the change of charts procedure, that is a finite (or
 infinitesimal) reparametrization  procedure which  modifies the frame
 leaving the physics unchanged.  This recipe defines so to speak a
 "geometrical stability" for all of kinds of symmetries which are
 locally realized.  This stability requirement must not be confused
 with the algebra of stability for the redefinitions of the physical
 parameters.   In this paper we discuss the former aspect in the framework
 of the  $\cW$ algebras \cite{Zamolodchikov:1985wn},
 realized over Riemann surfaces \cite{Gieres:1993sd}
-\cite{Bonora:1989wn}.  The intricacies  encountered  are well known
and can be found in the literature \cite{bol}
-\cite{mat}, and the outstanding mathematical books \cite{gunning} as well.

Even if the $\cW$ algebras arise from many sources
 \cite{KN}
 -\cite{Schlichenmaier:1990iv},
  the proper definition of all the $\cW$ 
 observables over the whole Riemann surface is not a simple affair. 
 For sake of truth this problem occurs in the usual conformal
 models \cite{Becchi:1988as}, 
 but in the context of $\cW$ algebras it becomes relevant\cite{Gervais:1982nw}
 -  \cite{Bakas:1989um}. The capital role payed by a $\cW$ symmetry has
  been pointed out in many physical fields ,
 such as integrable models \cite{Das:1991cn},\cite{Das:1991dc},
  string theory \cite{Friedan:1986ge} and supersymmetry\cite{Crane:1988uf}
 -   \cite{Delduc:1990gn}. Many  efforts have been made in that direction and 
 important results have been 
 reached \cite{radul}
-  \cite{Adler:1979ib}. The physical symmetry constituents are currents whose O.P.E (or different recipes),
 generate extensions of the conformal symmetry\cite{Bryant:1990sn}
 -  \cite{polya}:for this reason 
they are usually defined as "primary fields".
Their equations of motion define the dynamics of the system 
and, since they are  defined on a Riemann surface\cite{Bais:1988dc}
  -  \cite{Fateev:1987vh},
they must be  promoted to the status of geometrical objects. Since any
physical investigation cannot be separated from any geometrical
consideration, the same kind of symmetry arises in the context of
writting down welldefined differential equations on Riemann surfaces  
\cite{Gervais:1982nw}
- \cite{ly}.

The main difficulty lies in the definition of differential
 operators on Riemann surfaces \cite{bol}; in particular  their equipment
 in a ``conformal covariant" frame has been widely discussed in the literature
\cite{GustPee},\cite{gunning}\cite{matone}.                                         
In particular the more general question of classifying all
holomorphically covariant differentials operators amounts to the study
of operators which are  
covariant with respect to projective transformations.
At the end, the most general holomorphic covariant 
operators become a superposition of ordinary derivative operators 
with functions coefficients, whose reparametrization properties ensure the full covariantization of the operator.
The latter (or some of their combinations), within a  well defined mathematical framework,
become generators  of $\cW$-algebras 
\cite{Wang:1990np},\cite{diz1991}\cite{Mathieu:1988pm},\cite{Bakas:1989um},
\cite{Das:1991cn}.

A natural question which is now to be asked concerns the link between
these geometrical 
 coefficients and the physical ``primary fields" currents arising in $W$-symmetry.
There are other way of controlling the intricacies of covariance and 
associativity of the symmetry in $\cW$ geometry by the use of
appropriate holomorphic  
bundles, such as the jet bundles 
\cite{Zucchini1}, the Drinfeld-Sokolov bundle \cite{Zucchini2,Zucchini3},
 and the Toda bundle \cite{Aldrovandi-Bonora}.
However, we believe that $\cW$-algebras describe a local symmetry, for
this reason  
we have recently proposed \cite{symplecto,complexstruct,w3} an approach to $\cW$-algebras in a
symplectic framework. First, the 
chiral $\cW_\infty$-algebra \cite{Hull} emerges from the conspiration of
the infinitesimal action of symplectomorphisms together  
with suitable reparametrizations of the two dimensional local complex $\zbz
$ coordinates over a compact Riemann surface \cite{symplecto}. In this
approach the complex structure plays, as usual, 
an essential role \cite{complexstruct}, and the geometry of the
outstanding of finite $\cW$-algebras can be studied \cite{w3}.  

The $\cW$-algebras govern \cite{symplecto,complexstruct}
a hierarchy of smooth changes of local complex coordinates
 $\zbz\lra(\ZBZrz)$,   $ r=1,\cdots n$ 
from the $\zbz$ background local complex coordinates to a sequence
of suitable  $(\ZBZrz)$ ones through a symplectic scenario.
We shall follow a B.R.S approach in order to fully exploit the locality
of the theory. Accordingly, instead of considering non local commutators (or
Poisson brackets) between some primary fields, 
we shall analyse the B.R.S. transformations of the
Fadeev-Popov ghosts.  

The so-called chiral B.R.S. ghost \cite{Becchi:1988as}, $\cK^{(r)}$
associated to the 
hierarchy of smooth changes of complex coordinates  can be
 decomposed into a sum  of the other ghost fields $\cC^{(j)}\zbz$
pertaining to the infinite $\cW$-symmetry and whose 
coefficients are fixed by the geometrical space 
 content. In our scheme all the elements of the algebra acquire a
 well defined geometrical meaning,  and each of the ghosts
 $\cC^{(r)}\zbz$  behaves as a $(-r,0)$-differential.  So this
approach includes together both a hierarchy of smooth changes of
complex coordinates
 and the $\cW$ transformations.  We have pointed out in \cite{w3} that
 there are two different physical situations: the first one occurs when
 th hierarchy realizes a physical symmetry, the second
 one when this symmetry is broken, but the associativity requirements
 are preserved.  In this set of circumstances arise the primary
 fields, and the broken symmetry can be ruled by means of a constant
 Faddeev-Popov field $\theta$ which, notwithstanding  the breaking
 terms, secures the validity of the Jacobi identity \cite{Band JMP91}.
 So the whole symplectic space must be doubled in a $0$ component
(where the physics lives) and a $\theta$ partner (which guarantees the 
 algebra closure).  The coordinate transformations will
 be written as:

\begin{eqnarray}
&&\sS_{\cW} Z^{(r)}_0\zbz = \cK^{(r)}_0\zbz \prt Z^{(r)}_0\zbz
 +Z^{(r)}_\theta\zbz \nn\\[2mm] 
&&\sS_{\cW} Z^{(r)}_\theta\zbz =
 \cK^{(r)}_0\zbz \prt Z^{(r)}_\theta\zbz -  \cK^{(r)}_\theta\zbz \prt
 Z^{(r)}_\theta\zbz  \lbl{szr}
\lbl{sZZ}
\end{eqnarray}
 with the holomorphic ghosts transformations:
\begin{eqnarray}
&&\sS_{\cW}(\cK^{(r)}_0\zbz +\theta    \cK^{(r)}_\theta\zbz)\nn\\
&&\hskip 1cm =\biggl(\cK^{(r)}_0\zbz +\theta
\cK^{(r)}_\theta\zbz\biggr)\prt\biggl(\cK^{(r)}_0\zbz +\theta
\cK^{(r)}_\theta\zbz\biggr),
\end{eqnarray}
 and $\sS_{\cW}\theta=-1$.

In our symplectic formalism all the quantities acquire a well defined
geometric meaning. In particular, the holomorphic ghosts give an origin
to a decomposition which generate the finite $\cW$-algebra generators
$\cC^{(j)}\zbz$ and $\cX^{(j)}\zbz$, by setting
  
\begin{eqnarray}
\cK^{(r)}_0\zbz + \theta \cK^{(r)}_\theta\zbz &=&
\sum_{j=1}^r
\biggl(\omega^{(r)}_{(j-1),0}\zbz +\theta
\omega^{(r)}_{(j-1),\theta}\zbz\biggr)\\
&&\hskip 1cm  \biggl(\cC^{(j)}\zbz+\theta \cX^{(j)}\zbz\biggr) \lbl{kn}\nn
\end{eqnarray}
where the intertwining
 coefficients $\omega^{(r)}_{(j-1),0}\zbz$ and
 $\omega^{(r)}_{(j-1),\theta}\zbz$ 
 can be recursively expressed in terms of the coordinates as
\begin{eqnarray}              
&&\omega^{(r)}_{(j-1),0}\zbz   =
j! \prod_{i=1}^{m_j}
\biggl[
\frac{{(\prt  Z^{(p_i)}_0\zbz ) }^{a_i  }}{{a_i}!
 \prt  Z^{(r)}_0\zbz}
  \biggr]_{\bigg|_
{\tiny{\left\{
\begin{array}{c}
{\sum_i a_i =j},\
{\sum_i a_i p_i=r}\\[2mm]
{p_1>p_2>\cdots>p_{m_j}} 
\end{array}\right\}}} }
\nn\\
&&{}\\
&&\omega^{(r)}_{(j-1),\theta}\zbz= j! \prod_{i=1}^{m_j}
\biggl[
\frac{{
(  \prt  Z^{(p_i)}_\theta\zbz) }^{a_i  }}{{a_i}!
 \prt  Z^{(r)}_0\zbz} -
\frac{{(
  \prt  Z^{(p_i)}_0\zbz) }^{a_i  } \prt  Z^{(r)}_\theta\zbz }{{a_i}!
{( \prt  Z^{(r)}_0\zbz)}^2}   
  \biggr]_{\bigg|_
{\tiny{\left\{
\begin{array}{c}
{\sum_i a_i =j},\
{\sum_i a_i p_i=r}\\[2mm]
{p_1>p_2>\cdots>p_{m_j}} 
\end{array}\right\}}} }\nn
\end{eqnarray} 
and in turn are completely fixed by the geometry and behave as true tensors.
  
At the end the symplectic B.R.S. transformations 
imply that the variations of the 
$\cC^{(l)}\zbz$ ghost fields describe 
the chiral part of $\cW$-algebras \cite{symplecto,w3}:

\begin{eqnarray}
&&\sS_{\cW}\cC^{(l)}\zbz =\sum_{s=1}^l s\cC^{(s)}\zbz \prt \cC^{(l-s+1)}\zbz
+\cX^{(l)}\zbz, \qquad l=1,\dots, n,
 \lbl{scr} 
\end{eqnarray}
where the $\cX^{(l)}\zbz$ breaking terms represent, in the B.R.S approach, the
contribution to the finite $\cW$-algebras coming from the O.P.E. 
between primary fields,  
and indeed break the chiral representation invariance, but still preserve
associativity \cite{w3}. 
In the finite $\cW_3$ instance we have:
\begin{eqnarray}
\sS_{\cW_3} \cC^{(1)}\zbz&= &\cC^{(1)}\zbz\prt\cC^{(1)}\zbz
-
 \frac{16 }{3}\cT\zbz \cC^{(2)}\zbz\prt \cC^{(2)}\zbz   \nn\\
&&\qquad +\, \biggl( 
\prt\cC^{(2)}\zbz\prt^2\cC^{(2)}\zbz 
- \frac{2}{3}\cC^{(2)}\zbz\prt^3\cC^{(2)}\zbz\biggr)
\lbl{sc1} \\
\sS_{\cW_3}
\cC^{(2)}\zbz&=&\cC^{(1)}\zbz\prt\cC^{(2)}\zbz+2\cC^{(2)}\zbz\prt\cC^{(1)}\zbz
\nn 
\end{eqnarray}
In order to insure the nilpotency of the BRS operation,
$\cX^{(l)}\zbz$ in Eq\rf{scr} must transform as:
\begin{eqnarray}
\sS_{\cW} \cX^{(l)}\zbz&=& \sum_{s=1}^l \biggl(
 s\cC^{(s)}\zbz \prt\cX^{(l-r+1)}\zbz\biggr)- s\cX^{(s)}\zbz \prt
 \cC^{(l-s+1)}\zbz.
\lbl{scr1}
\end{eqnarray}
The full completion of the last equation \rf{scr1} amounts to
introducing together with thei $\cW$-variations) all the set of the
primary fields (including  
fields not appearing in $\cX^{(l)}\zbz$ ) which belong to the
algebra. 
In the $\cW_3$ case we obtain:
\begin{eqnarray}
\sS_{\cW_3} \cT\zbz &=&\cC\zbz\prt\cT\zbz +2 \cT\zbz\prt\cC\zbz 
{-}
\cW\zbz \prt\cC^{(2)}\zbz\nn\\
&{-}&
\frac{2}{3}
\cC^{(2)}\zbz 
\prt\cW\zbz +\prt^3 \cC\zbz ,
\lbl{ct1}
\end{eqnarray}
 which force to introduce a cubic differential $\cW\zbz$ as a field
 not involved in 
 $\cX^{(1)}\zbz$; the minimal nilpotency chain closes by:   
\begin{eqnarray}
\sS_{\cW_3} \cW\zbz &=&\cC\zbz \prt \cW \zbz+ 3\prt\cC\zbz \cW\zbz
 + 
16
\cT\zbz
\prt\biggl(\cC^{(2)}\zbz\cT\zbz\biggr)\nn\\
&+&\biggl(\prt^5\cC^{(2)}\zbz
+
2
\cC^{(2)}\zbz\prt^3\cT\zbz
+10\cT\zbz\prt^3\cC^{(2)}\zbz\nn\\
&+&15\prt\cT\zbz\prt^2\cC^{(2)}\zbz
+
9
\prt^2\cT\zbz\prt\cC^{(2)}\zbz\biggr)
\lbl{scw}
\end{eqnarray}
The  structure of Eqs\rf{scr}\rf{scr1}  fixes a hierarchy such that
the lowest orders are fixed by the higher ones; 
so the highest breaking term $\cX^{(n)}\zbz$ 
is so to speak related to the lowest order terms.

In the paper we aim to study the properties of the
$\cX^{(l)}\zbz$  quantities under finite and infinitesimal change of
charts in order to  guarantee a global definition of the finite $\cW$-algebra  
over a Riemann surface.
Our investigation will lead to the result that primary fields are related
to the function coefficients required to build up the conformal covariant
derivatives and as such they describe gravitational degrees of freedom.
This will be shown first in Section 2 through a 
finite change of charts. Anyhow this procedure does not fully exploit
the locality values.
To utilize them we further use the infinitesimal  local change of
charts in a B.R.S. way. 
The derived consistency requirements in this approach 
generate a connection between functions containing primary
fields and the coefficients necessary to define the action of the
holomorphic derivative in an intrinsic way.  
This link proves the inverse of the result, previously cited, found by Refs 
\cite{Wang:1990np},\cite{diz1991}\cite{Mathieu:1988pm},\cite{Bakas:1989um},
\cite{Das:1991cn} , and it is obtained  within a fully  general
covariance calculations. 
As a concluding remark we stress that this result greatly takes some
benefits from the local approach to $\cW$ algebras we have recently provided
in Refs \cite{symplecto,complexstruct,w3}. 
 
\sect{The stability of $\cW$-algebra under holomorphic change of charts }

The purpose of this Section is to investigate the consequences of the
background holomorphic change of coordinates
\begin{eqnarray}
\wbw\lra \zbz,\quad w=w(z)
\lbl{changeframe}
\end{eqnarray}
for the two equations \rf{scr}\rf{scr1}. 
To hold its validity over the whole surface, each member of each
equation must verify the same transformation law. 
It is obvious that changes of charts
and $\cW$-symmetry will commute. 
Since the fields $\cC^{(l)}\zbz$ behave as $(-r,0)$-differentials

\begin{eqnarray}
&&\cC^{(l)}\wbw={(w')}^l\cC^{(l)}\zbz\nn\\
\lbl{changeC}
\end{eqnarray} 
each term of Eqs \rf{scr} and \rf{scr1} must transform in the same way.

Thanks to the ghost property, a simple calculation shows that the
chiral summand 
$$\sum_{s=1}^l s\cC^{(s)}\zbz \prt \cC^{(l-s+1)}\zbz$$   
behaves under finite change of charts as a $(-l,0)$-differential, so that
problems come from  
the presence of the $\cX^{(l)}\zbz$ breaking term which by
covariance must transform also as a $(-l,0)$-differential

\begin{eqnarray}
&&\cX^{(l)}\wbw={(w')}^l\cX^{(l)}\zbz . 
\lbl{changeX}
\end{eqnarray}
According to the Faddeev-Popov grading and for $l=1,\dots,n$ the
,$\cX^{(l)}$ breaking term is decomposed
over the ghost monomials but $\cC^{(1)}$ as
\begin{eqnarray}
&&\cX^{(l)}\zbz = \frac{1}{2}\sum_{r,s\geq 0}\sum_{p,q =2}^n \prt^r
\cC^{(p)}\zbz\prt^s \cC^{(q)}\zbz
\Lambda^{(l)}_{p+q-r-s-l}(r,p|s,q|l)\zbz,
\lbl{Lbda}
\end{eqnarray}
where the zero graded coefficient functions
$\Lambda^{(l)}_{p+q-r-s-l}(r,p|s,q|l)$ is skewsymmetric under the
permutation of the pairs $(r,p)\leftrightarrow(s,q)$ and contain
primary fields.

\subsection{Behaviour under finite holomorphic change of complex coordinates}

As is well known troubles come from the derivative operators: in fact
the $r$-th derivative of a $p$-th order  tensor is no more a tensor
and  transforms in a convoluted way. 
For this reason we shall use `normalized' covariant Bol operators $L_r$
\cite{Gieres:1993sd} \cite{bol}  :   
\begin{eqnarray}
&&L_r\zbz =\sum_{j=0}^r a^{(r)}_j\zbz \prt^{r-j},\qquad
a^{(r)}_0\zbz=1,\qquad
a^{(r)}_1\zbz=0
\lbl{lr}
\end{eqnarray}
which under any change of holomorphic charts transform as

\begin{eqnarray}
&&L_r\wbw={(w')}^{-\frac{1+r}{2}}L_r\zbz{(w')}^{\frac{1-r}{2}}.
\lbl{boltransform}
\end{eqnarray} 
$L_r$ transforms functions of conformal weight $\frac{1-r}{2} $
 into ones of weight $\frac{1+r}{2} $.
In particular the kernel of $L_r\wbw$  
is $r$ dimensional linear space which is stable under any holomorphic
 change of charts $\wbw \lra \zbz$. 
 
This covariance property enables a (non unique) construction
 of the $ a^{(r)}_j\zbz $ coefficients \cite{diz1991} within a coordinate description.
How this construction can be carried out in our coordinate scheme is
 discussed in the Appendix; 

The explicit expressions in terms of the coordinates 
for the coefficients $a_2^{(2)}\zbz $,  $a_2^{(3)}\zbz $ and
$a_3^{(3)}\zbz $ (the latter will be useful to discuss the $\cW_3$-algebra)
are: 

\begin{eqnarray}
a_2^{(2)}\zbz &=& \prt^2ln \prt Z\zbz  -\frac{1}{2}{(\prt \ln \prt Z\zbz )}^2\nn \\
a_2^{(3)}\zbz  &=& \prt^2 \ln w\zbz  - \frac{1}{3} (\prt \ln w\zbz )^2 +
\frac{v\zbz }{w\zbz }\\
a_3^{(3)}\zbz  &=& \frac{1}{3} \prt^3 \ln w \zbz
 + \frac{2}{27} (\prt \ln w\zbz )^3 
+ \frac{v\zbz }{3w\zbz } \prt \ln w\zbz \nn
\lbl{avw}
\end{eqnarray}

where we have set the determinants 
\begin{eqnarray}
w \zbz = \left| \begin{array}{cc} 
\prt Z\zbz  & \prt Z^{(2)}\zbz \\
\prt^2 Z \zbz & \prt^2 Z^{(2)} \zbz  
\end{array} \right|, \quad 
v\zbz = \left| \begin{array}{cc} 
\prt^2 Z \zbz & \prt^2 Z^{(2)}\zbz \\
\prt^3 Z \zbz & \prt^3 Z^{(2)}\zbz  
\end{array} \right|.
\lbl{determinants}
\end{eqnarray}

Turning back to the glueing problem of the breaking term $\cX^{(l)}$,
without any loss of generality, \rf{Lbda} rewrites for $l=1,\dots,n$ 
\begin{eqnarray}
&&\cX^{(l)}\zbz = \frac{1}{2} \sum_{r,s\geq 0}\sum_{p,q= 2}^n
L_r\cC^{(p)}\zbz L_s\cC^{(q)}\zbz\cA_{p+q-r-s-l}^{(l)}(r,p|s,q|l)\zbz,
\lbl{cA}
\end{eqnarray}
where $\cA_{p+q-r-s-l}^{(l)}(r,p|s,q|l)$ is of zero ghost grading,
skewsymmetric under the  permutation of the pairs
$(r,p)\leftrightarrow(s,q)$ and constructed over some primary fields.
 
It is easy to see that  the derivative troubles can be overcome by inverting
 equation \rf{lr} iteratively,

\begin{eqnarray}
&&\prt^r =\sum_{j=0}^r b^{(r)}_j\zbz L_{r-j}\zbz,\qquad
b^{(r)}_0\zbz=1,\quad
 b^{(r)}_1\zbz=0,
\lbl{derivatives}
\end{eqnarray}
and where each of the $b^{(r)}_j$ for $j\geq 2$ is a polynomial
expression in the $a^{(s)}_k$, namely for
\begin{eqnarray}
j\geq 2,\quad  J_k := \sum_{m=0}^{k-1} j_m,\quad J_1 =0, \quad
b^{(r)}_j = \sum_{i=1}^r (-1)^i \left[ \prod_{\ell=1}^{i}
\left( \sum_{j_\ell =1}^{r-J_\ell} a^{(r-J_\ell)}_{j_\ell} \right)
\right]\delta_{j,J_{i+1}}. 
\lbl{ba}
\end{eqnarray}
Equation \rf{changeX} thus writes:
\begin{eqnarray}
\lefteqn{\sum_{r,s\geq 0}\ \sum_{p,q= 2}^n
L_r\wbw\cC^{(p)}\wbw
L_s\wbw\cC^{(q)}\wbw\cA_{p+q-r-s-l}^{(l)}(r,p|s,q|l)\wbw}\nn\\ 
&&={(w')}^l\sum_{r,s\geq 0}\ \sum_{p,q= 2}^n
L_r\zbz\cC^{(p)}\zbz
L_s\zbz\cC^{(q)}\zbz\cA_{p+q-r-s-l}^{(l)}(r,p|s,q|l)\zbz\nn\\  
&&\equiv
\sum_{r,s\geq 0}\ \sum_{p,q= 2}^n
 {(w')}^{-\frac{(1+r)}{2}}\Biggl[L_r\zbz{(w')}^{\frac{(1-r)}{2}+p}\cC^{(p)}\zbz\Biggr]\lbl{ltensor}\\  
&& {(w')}^{-\frac{(1+s)}{2}}\Biggl[L_s\zbz{(w')}^{\frac{(1-s)}{2}+q}\cC^{(q)}\zbz\Biggr]\cA_{p+q-r-s-l}^{(l)}(r,p|s,q|l)\wbw\nn  
\end{eqnarray}

Using successively \rf{lr}, Leibniz
rule and \rf{derivatives} one gets after some combinatoric manipulations

\begin{eqnarray}
{(w')}^{-\frac{1+r}{2}} L_r\zbz\Biggl[{(w')}^{(\frac{1-r}{2}+p)}
\cC^{(p)}\zbz\Biggr] = 
\sum_{m=0}^r \alpha(r,m,p)\zbz L_m\cC^{(p)}\zbz,
\lbl{LinL}
\end{eqnarray}
with the polynomial in the $a^{(s)}_k$
\begin{eqnarray}
\alpha(r,m,p)\zbz = \sum_{j=0}^{r-m} \sum_{k=m}^{r-j}
\left(\begin{array}{c}r-j\\k\end{array}\right) 
{(w')}^{-\frac{1+r}{2}} a^{(r)}_j \zbz b^{(k)}_m\zbz 
\prt^{r-j-k} {(w')}^{\frac{1-r}{2}+p}.
\end{eqnarray}
Inserting twice \rf{LinL} into \rf{ltensor} we finally deduce 
a set of algebraic equations 

\begin{eqnarray}
\cA_{p+q-r-s-l}^{(l)}(r,p|s,q|l)\zbz &=& \sum_{m\geq r} \sum_{m'\geq s}
w'^{-l}\alpha(m,r,p)\zbz \alpha(m',s,q)\zbz\nn\\
&&\hskip 2cm 
\cA_{p+q-m-m'-l}^{(l)}(m,p|m',q|l)\wbw 
\lbl{finchangeofcharts}
\end{eqnarray}
In particular, this system gives for each value of $ p,q,r,s,l$
the glueing properties of the function $\cA_{p+q-r-s-l}^{(l)}(r,p|s,q|l)$.
Restricting the system \rf{finchangeofcharts} to the scalar sector,
that is for $p+q-r-s-l=0$ and 
$$ 
\cA_{0}^{(l)}(r,p|s,q|l)\zbz\arrowvert_{r+s-p-q-l=0} =
\cA_{0}^{(l)}(r,p|s,q|l)\wbw\arrowvert_{r+s-p-q-l=0}
$$

the scalar component of $\cA^{(l)}_{p+q-r-s-l}(r,p|s,q|l)$ cancels out
and the system reduces to an homogeneous one containing
$\cA^{(l)}_{r+s-m-m'}(m,p|m',q|l)$, $m+m'>r+s$ with therefore negative
lower indices. The  finite degree $n$ of the algebra guarantees a
finite number of equations $l=1,\dots,n$, and the determinant of the
system will be non zero in general leading to
$\cA^{(l)}_h \equiv 0$ for $h<0$ and $l=1,\dots,n$.
We reach the conclusion:

\begin{theorem}
 The stability under change of charts  requires that the only non
vanishing  
$\cA^{(l)}_{p+q-r-s-l}(r,p|s,q|l)$ have a positive
lower indices content, $p+q-r-s-l\geq 0$. 
\lbl{changeofchartstheorem}
\end{theorem}

Anyhow in the more general cases, the way out to get fully local relations in only one argument, is to use the 
infinitesimal holomorphic change of coordinates. 

\subsection{Infinitesimal holomorphic change of complex coordinates}

Under infinitesimal holomorphic change of coordinates
 $z\longrightarrow z - \epsilon(z)$ and following the notation used in
 \cite{Gieres:1993sd} we have :

\begin{eqnarray}
&&\delta_\epsilon \cC^{(m)}\zbz =X_{-m}(z)\cC^{(m)}\zbz ,
\qquad
X_{m}(z)=\epsilon(z)\prt +m\prt\epsilon (z)
\end{eqnarray}
At the infinitesimal level, \rf{changeX} reduces to
\begin{eqnarray}
\delta_\epsilon   \cX^{(l)}\zbz   =X_{-l}(z)\cX^{(l)}\zbz 
\lbl{deltaX}
\end{eqnarray} 
while the conformal covariant derivative 
transforms as \cite{Gieres:1993sd,diz1991} 

\begin{eqnarray}
&&\Biggl[\delta_\epsilon, L_r\Biggr] =X_{\frac{(1+r)}{2}}L_r -L_r
X_{\frac{(1-r)}{2}} 
\lbl{commepsilon}
\end{eqnarray}
and in particular the case $r=1$ gives $[\delta_\epsilon, \prt]=0$.
Accordingly the $a_j^{(k)}$ coefficients satisfy
\begin{eqnarray}
\delta_\epsilon  a_j^{(k)}\zbz &=&
X_j(z) a_j^{(k)}\zbz +\frac{1}{2}
\left(\begin{array}{c}k+1\\k-j\end{array}\right)(j-1)\prt^{j+1}\epsilon(z) 
\cr
&&-\ \sum_{l=2}^{j-1}\Biggl\{
\left(\begin{array}{c}k-l\\j-l+1\end{array}\right)-\frac{k-1}{2}  
 \left(\begin{array}{c}k-l\\j-l\end{array}\right)\Biggr\}a_l^{(k)}\zbz
\prt^{j-l+1}\epsilon(z)  
\lbl{sa}
\end{eqnarray}
for $j\geq 2$. Note that \rf{sa} is linear in $a_j^{(k)}$ and depends
on its first derevative only. Turning the holomorphic
$(-1,0)$-differential 
$\epsilon(z)$ a Faddeev-Popov ghost (and accordingly $X_l$ as well),
so that $\delta_\epsilon^2=0$, one has :
\begin{eqnarray}
\delta_\epsilon \epsilon(z)=\frac{1}{2}X_{-1}\epsilon(z)=
\epsilon(z)\prt\epsilon(z) 
\end{eqnarray}
and using \rf{commepsilon} for $r=1$, one finds
\begin{eqnarray}
\biggl\{\delta_\epsilon,X_l(z)\biggr\}=\epsilon(z)\biggl[\prt,X_l(z)\biggr]
\lbl{commX}
\end{eqnarray} 
so that the operator $\delta_\epsilon -X_l$ is nilpotent for any $l$
in turn.

The independence of the tranformations assure that the $\cW$-tranformations
 anticommute with the $\delta_{\epsilon}$ operator:
\begin{eqnarray}
\sS_{\cW}\epsilon(z)=0,\qquad \{\delta_\epsilon,\sS_{\cW}\} = 0.
\lbl{anticom}
\end{eqnarray}

It could be noted that the underlying algebra of $\delta_{\epsilon}$
is equivalent to the $\cW$-algebra where the chiral ghosts
$\cC^{(l)}\zbz\quad l\geq 2$ are set to be zero which is nothing but
the usual conformal algebra found in \cite{Becchi:1988as}.
In our approach the $\cW$-algebras come out from the action of smooth cotangent
diffeomorphisms on a chain of smooth changes of local complex
coordinates on the Riemann surface.  In turn, all the $\ZBZrz$
complex coordinates have well defined $\cW$-transformations, and the B.R.S.
algebra is kept nilpotent by performing a $\theta$ doubling in order
to obey the Jacobi identities.  This doubling trick amounts to
studying the true structure of the $Z^{(r)}_\theta\zbz$
coordinates. The question will be partially  solved (this will be
enough for the purpose) by 
looking at the properties of the infinitesimal change of charts of
$Z^{(r)}_0\zbz$ and their  B.R.S.  $\cW$-transformations as well.

The $Z^{(r)}_0$ complex coordinates are scalars under
change of charts:  
\begin{eqnarray}
&&\delta_\epsilon Z^{(r)}_0\zbz =\epsilon(z)\prt Z^{(r)}_0 \zbz
\end{eqnarray}
while in their $\cW$-transformations,
see \rf{sZZ}, the most general form of $Z^{(r)}_\theta$ can been made
 explicit in terms of the chiral ghosts \cite{w3}, 
\begin{eqnarray}
\cS_{\cW} Z^{(r)}_0\zbz = \cK^{(r)}_0\zbz \prt Z^{(r)}_0\zbz
+\sum_{s\geq 0}\sum_{p=2}^n L_s \zbz\cC^{(p)}\zbz
\cB^{(r)}_{p-s}(p|s|r)\zbz .
\end{eqnarray}
The anticommutativity \rf{anticom} between the two operations on
$Z^{(r)}_0$ and the use of \rf{commepsilon} yield
\begin{eqnarray}
(\delta_\epsilon - X_{p-s}(z))\cB^{(r)}_{p-s}(p|s|r)\zbz = 
\cN^{(r)}_{p-s}(p|s|r|a,\cB)\zbz
\end{eqnarray}
where $\cN^{(r)}_{p-s}$ is linear in the $\cB$s
and of grading one with respect to $\varepsilon$ of the form
\begin{eqnarray}
\cN^{(r)}_{p-s}(p|s|r|a,\cB)\zbz = \sum_{k\geq 2}
\prt^k\epsilon(z) \sum_{m\geq s+k+1} N_{m+1-k-s}(p|s|k|m;a)\zbz
\cB^{(r)}_{p-m}\zbz
\lbl{cn}
\end{eqnarray}
where $N_{m+1-k-s}(p|s|k|m;a)$ is a known polynomial in the $a^{(s)}_k$
coefficients,
\begin{eqnarray}
N_{m+1-k-s}(p|s|k|m;a) = \sum_{j=0}^{m+1-s-k} \!\!
\left(\frac{1-m}{2} + p\right)\!
\left(\!\begin{array}{c} m-j\\ m+1-k-j\end{array}\!\right)
a^{(m)}_j\, b^{(m+1-k-j)}_{m+1-k-j-s}.
\lbl{defN}
\end{eqnarray}
 By nilpotency, one gets
\begin{eqnarray}
(  \delta_\epsilon - X_{p-s} )\cN^{(r)}_{p-s}(p|s|r|a,\cB)=0.
  \end{eqnarray} 
Owing to \rf{sa} this expression is still
algebraic in the coefficients $a^{(s)}_k$ and their first order
$z$-derivatives. 
Decomposing over the monomials $\prt^k\epsilon\prt^\ell\epsilon$ 
leads to a homogeneous linear system in the $\cB$s that can be
inverted. Hence expressing algebraically the $\cB$s in terms of the
$a^{(s)}_k$ coefficients and their first order
$z$-derivatives
implies that $Z^{(s)}_\theta$ is expressible by means of both $\Zr_0$
and $\cC^{(\ell)}$ and their $z$-derivatives, see the appendix. 
  
Turning back to the study of the behaviour of $\cA$ under an
infinitesimal holomorphic change of complex coordinates.
Let us now derive the properties of 
$\cA_{p+q-r-s-l}^{(l)}(r,p|s,q|l)$ from \rf{cA} and \rf{deltaX}.
After some calculation we get:
\begin{eqnarray}
\bigg(\delta_\epsilon - X_{p+q-r-s-l}(z)\bigg) 
\cA_{p+q-r-s-l}^{(l)}(r,p|s,q|l)\zbz
=\Omega_{p+q-r-s-l}^{(l)}(r,p|s,q|l)\zbz
\lbl{sA}
\end{eqnarray}
where $\Omega_{p+q-r-s-l}^{(l)}(r,p|s,q|l)$ is the following linear
expression in the $\cA^{(l)}$ for 

 \begin{eqnarray}
&& p,q\geq 0,\quad p,q = 2,\dots,n,\quad l=1,\dots,n\nn\\[4mm]
\Omega_{p+q-r-s-l}^{(l)}(r,p|s,q|l)\zbz &= & \sum_{k\geq 2}
\prt^k\epsilon(z)\times \cr 
&& \hskip -4cm
\left(\sum_{m\geq r+k+1} \!\! N_{m+1-k-r}(p|r|k|m;a)\zbz
\cA^{(l)}_{p+q-m-s-l}(m,p|s,q|l)\zbz \right. \cr
&&\hskip -4cm
+ \left. \sum_{m\geq s+k+1} N_{m+1-k-s}(p|s|k|m;a)\zbz
\cA^{(l)}_{p+q-r-m-l}(r,p|m,q|l)\zbz \right),
 \end{eqnarray}
where $N$ has already been defined in \rf{defN}. 
 
Once more by nilpotency, one has
\begin{eqnarray}
\bigg(\delta^2_\epsilon - X_{p+q-r-m-l}(z) \bigg)
\Omega_{p+q-r-s-l}^{(l)}(r,p|s,q|l)\zbz = 0.
\lbl{deltaOmega}
\end{eqnarray}
Solving this condition insures the proper definiteness of the whole
theory over a Riemann surface. 
As before decomposing over the monomials $\prt^k\epsilon\prt^\ell\epsilon$
leads for each sector a homogeneous linear system in the $\cA$s in
finite dimension thanks to Theorem \rf{changeofchartstheorem} and the
finite degree of $\cW$-algebra that can be inverted.

This shows that trivial solutions $\cA^{(\ell)}$ of Eq\rf{deltaOmega}
are algebraic expressions in the coefficients $a^{(l)}_j\zbz$ and
their first order $z$-derivatives.
Therefore they are related to the complex coordinates $\Zr_0$.
This drastically changes the physical scenario, namely, if we rely on
the idea that primary fields describe a conformal matter we have to
change our mind and relate them to gravitational degrees of freedom. 
 
The closure under of the B.R.S. $\cW$-tranformations of the  
$\Zr_0$  fields secures that all of the necessary primary fields,
not involved in the breaking terms $\cX^{(l)}$,
will be also differential polynomials (with higher order
$z$-derivatives) in the coefficients $a^{(l)}_j$. They accordingly
turn to be also functions of the $\Zrz$ coefficients and their derivatives.
Collecting together these results yields the Statement:
\begin{Ansatz}
All primary fields involved in a given finite $W_n$-algebra are local
functions of the $\Zr_0$ complex coordinates and their $z$-derivatives. 
\lbl{Ansatz}   
\end{Ansatz} 
The latter completely modifies the physical point of view of this scenario. 

The identifications of the old primary fields in terms of monomials
 of $a^{(l)}_j\zbz$ and their derivatives was performed in Ref \cite{diz1991}: our treatment gives the exactly  inverse of the  proof given there:
 while in this reference it was proved that the $a^{(l)}_j$ can be
expressed in terms of the generators $w$ of 
a given $\cW$-algebra. Here,  we state the converse, namely,
 that the primary fields of this algebra
can be expressed with the  $a^{(l)}_j$ and their $z$-derivatives. 
For the $W_3$ instance, the currents are related to the $a^{(l)}_j\zbz$
coefficients from the solutions of the equation\cite{diz1991}
 \begin{eqnarray}
 &&L_3 f_i\zbz=0\quad i=1,2,3\nn\\
 &&\sum_{i,j,k}\epsilon^{i,j,k}\prt^2 f_i\zbz\prt f_j\zbz f_k\zbz =0
 \lbl{l3}
 \end{eqnarray} 
by
\begin{eqnarray}
\cT\zbz = \frac{a_2^{(3)}\zbz }{2}, \quad \mbox{ and }\quad  \cW\zbz =
\frac{1}{8}\Big(\frac{1}{2} \prt a_2^{(3)}\zbz  - a_3^{(3)}\zbz \Big).
\end{eqnarray}
With the help of \rf{avw} the cubic differential $\cW$
current reads
\begin{eqnarray}
\cW\zbz &=& \frac{1}{24} \biggl(\frac{1}{2} \prt^3 \ln w\zbz 
 - \prt^2 \ln w \zbz  
\prt \ln w\zbz  - \frac{2}{9} (\prt \ln w\zbz )^3\biggr)\nn\\
& +& 
  \frac{1}{16} \biggl( \frac{\prt v\zbz }{w\zbz }
  - \frac{5v\zbz }{3w\zbz }\prt \ln w\zbz  \biggr)
\end{eqnarray}
and is nothing but a Laguerre invariant while 
$a_2^{(3)}\zbz $ and the combination
$9(\prt a_2^{(3)}\zbz  - 3 a_3^{(3)}\zbz )$ are both the so-called Painlev\'e
invariants \cite{Forsyth} as pointed out in \cite{Govin95}.

By the way from Eq\rf{l3}\rf{ct1}\rf{scw} is easy to derive\cite{Grimm}:
\begin{eqnarray}
&&\sS f_i\zbz =\cC\zbz\partial f_i\zbz - f_i\zbz \partial\cC\zbz + 2\Ch\zbz \partial^2 f_i\zbz - \partial\Ch\partial f_i\zbz\nn\\&& +
\frac{1}{3} f_i\zbz \partial^2\Ch\zbz + \frac{8}{3}\cT\zbz f_i\zbz \Ch\zbz.
\end{eqnarray}
 The coordinates are defined \cite{Ovsienko} as the scalars obtained from the previous solutions; for example
  $Z_0=f_2/f_1$ and $Z_0^{(2)}=f_3/f_1$. By
quotients the variations of the former turn out to be
\begin{eqnarray}
\sS Z_0\zbz &=& \cC\partial Z_0\zbz + 2\Ch\zbz \partial^2 Z_0\zbz \nn\\&-& 
\partial\Ch\zbz \partial Z_0\zbz - \frac{4}{3} \Ch\zbz 
\partial Z_0\zbz \partial\ln \omega\zbz\\
\sS Z^{(2)}_0 \zbz&=& \cC\zbz\partial Z_0^{(2)}\zbz + 2\Ch\zbz \partial^2 Z_0^{(2)}\zbz \nn\\
& -& \partial \Ch\zbz \partial
Z_0^{(2)} \zbz- \frac{4}{3} \Ch\zbz \partial Z_0^{(2)}\zbz \partial\ln \omega\zbz 
\end{eqnarray}
where $\omega\zbz$ has been introduced in Eq \rf{determinants}

\section{Conclusions}

Our results show how general principles may clarify the physical
landscape, in the sense that
the quantum extension of models which satisfy $\cW$-symmetry generated
by primary fields, should help the quantum description  of
gravitational degrees of freedom. 
This extension greatly complicates this research topic, but, on an
other hand, is extremely provocative. 

Also the Laguerre-Forsyth construction offers the possibility of
expressing primary current fields directly in terms of ``matter
fields" considered as solutions of linear differential equations with
algebraic integrals of the type $L\,f=0$ where $L$ is the most general
covariant operators. This type of equations has been previously
obtained in a systematic way \cite{Grimm} through vanishing curvature
conditions (a useful trick already used in
\cite{Bilal:1991wn,Zucchini1}). Even if there exist few Lagrangian models
giving rise to field equations containing Bol's operators
\cite{Zucchini1}, a general construction of such Lagrangians is still missing. 
Furthermore, the intimate link between these 
solutions and a hierarchy of complex coordinates $\Zr_0$ coming from
the symplectic 
scenario raises the question whether which of either coordinates or matter
fields are the most important at the physical level. Expressing
$\Zr_0$ in terms of the solutions of $L_n\,f=0$ as in \rf{Zf}
generates the breaking from $W_\infty$ to finite $W_n$-algebras where
$n$ is exactly the order of the generalized Bol operator $L_n$ and may
give a partial answer to this question. Note also that the
$\theta$-`trick' used in \cite{w3} for introducing the chiral breaking
terms $\cX$ seems to be in relation with the restriction to $\ker\,L_n$.

\section{Appendix }

The aim of this appendix is to 
give a coordinate description of the coefficients 
$ a^{(s)}_j\zbz \forall s=2\cdots n$:
in terms of the coordinates $\Zrz $, $r=1\cdots n$.

To do this we use the Di Francesco-Itzykson-Zuber \cite{diz1991} within our construction \cite{symplecto}.

Consider the space$\cV_{\frac{1-s}{2}}$ of the functions $f_i\zbz$  with holomorphic weigth $d=\frac{1-s}{2}$  
 solutions of the equation:
\begin{eqnarray}
L_{s} f_i\zbz =0 \quad i= 1\cdots s
\lbl{solutions}
\end{eqnarray}
such that:
\begin{eqnarray}
\left|\begin{array}{ccc}
\prt^{(s-1)}f_1\zbz&\cdots&\prt^{(s-1)}f_{(s)}\zbz \\
\vdots &\ddots& \vdots \\
\prt f_1\zbz&\cdots&\prt f_{(s)}\zbz\\ 
f_1\zbz&\cdots&f_{(s)}\zbz
\end{array}
\right|         =1
\end{eqnarray}

So for a whatever other function $f\zbz\in \cV_{\frac{1-s}{2}}$
the action $L_s f\zbz$ can be defined by:
\begin{eqnarray}
L_s f\zbz =
\left|\begin{array}{cccc}
\prt^{(s)} f\zbz&\prt^{(s)}f_1\zbz&\cdots&\prt^{(s)}f_{(s)}\zbz \\ 
\prt^{(s-1)}f\zbz&\prt^{(s-1)}f_1\zbz&\cdots&\prt^{(s-1)}f_{(s)}\zbz \\
\vdots&\vdots &\ddots& \vdots \\
\prt f\zbz&\prt f_1\zbz&\cdots&\prt f_{(s)}\zbz\\ 
f\zbz&f_1\zbz&\cdots&f_{(s)}\zbz
\end{array}
\right|         
\end{eqnarray}  
which is a quantity with weight $d=-\frac{1-s}{2}  $. 
So the $a^{(s)}_j\zbz$ can be identified with minors of this determinant. 

\begin{eqnarray}
a^{(s)}_j\zbz =
\left|\begin{array}{cccc}
\prt^{(s)} f\zbz&\prt^{(s)}f_1\zbz&\cdots&\prt^{(s)}f_{(s)}\zbz \\
\vdots&\vdots &\ddots& \vdots \\ 
\prt^{(j+1)}f\zbz&\prt^{(j+1)}f_1\zbz&\cdots&\prt^{(j+1)}f_{(s)}\zbz \\ 
\prt^{(j-1)}f\zbz&\prt^{(j-1)}f_1\zbz&\cdots&\prt^{(j-1)}f_{(s)}\zbz \\
\vdots&\vdots &\ddots& \vdots \\
\prt f\zbz&\prt f_1\zbz&\cdots&\prt f_{(s)}\zbz\\ 
f\zbz&f_1\zbz&\cdots&f_{(s)}\zbz
\end{array}
\right|         
 \lbl{coefficients}
\end{eqnarray}  
From the $f_i\zbz\quad i=1\cdots s$ functions we can define $s-1$ scalars : 
  
\begin{eqnarray}
Z^{(j-1)}\zbz =\frac{f_j\zbz}{f_1\zbz},\qquad
j=2\cdots s
\lbl{Zf}
\end{eqnarray}

which provide a local  $s-1$ dimensional system of coordinates.
Assuming that this system coincides with the coordinates previously
introduced we invert the procedure to construct a system of $f_i\zbz$ functions:  

First we define the quantity:

\begin{eqnarray}
\omega\zbz =\left|\begin{array}{ccc}
\prt^{(s-1)} Z^{(1)}\zbz&\cdots&\prt^{(s-1)} Z^{(s-1)}\zbz\\ 
\vdots &\ddots& \vdots \\
\prt Z^{(1)}\zbz&\cdots&\prt Z^{(s-1)}\zbz 
\end{array}
\right|
\end{eqnarray} 
is an object with weight $\frac{s(s-1)}{2}$,
so we can  build the $s$  linearly independent functions  with weight $d=\frac{1-s}{2}$  \cite{Ovsienko}
\begin{eqnarray}
 f_1\zbz &=&\omega^{-\frac{1}{s}}\zbz\nn\\
 f_2\zbz &=& Z^{(1)}\zbz \omega^{-\frac{1}{s}}\zbz\nn\\
& \vdots&\nn\\
f_s\zbz &=& Z^{(s-1)}\zbz   \omega^{-\frac{1}{s}}\zbz \nn\\
\end{eqnarray}
which provide a system of solutions for the
 Eq\rf{solutions}, with the coefficients
  $a^{(s)}_j\zbz$ given in Eq\rf{coefficients}.

\end{document}